# Tensor Charges, Quark Anomalous Magnetic Moments And Baryons.

## MEKHFI.M

*Physics department, Es-Senia University 31100 Oran-Algeria*
*And*
*International Center for theoretical physics Trieste.Italy*

**Abstract.** We propose an "ultimate" upgrade of the Karl- Sehgal (KS) formula which relates baryon magnetic moments to the spin structure of constituent quarks, by adding anomalous magnetic moments of quarks. We first argue that relativistic nature of quarks inside baryons requires introduction of two kinds of magnetisms, one axial and the other tensoriel. The first one is associated with integrated quark helicity distributions $\Delta_i - \Delta_{\bar{i}}$ (standard) and the second with integrated transversity distributions $\delta_i - \delta_{\bar{i}}$. The weight of each contribution is controlled by the combination of two parameters, $x_i$ the ratio of the quark mass to the average kinetic energy and $a_i$ the quark anomalous magnetic moment. The quark anomalous magnetic moment is correlated to transversity and both are necessary ingredients in describing relativistic quarks. The proposed formula, when confronted with baryon magnetic moments data with reasonable inputs, confirms that anomalous magnetic moments of quarks are unavoidable intrinsic properties.

# I- ANOMALOUS MAGNETIC MOMENTS OF QUARKS ARE NOT NEGLIGIBLE

Non relativistic constituent quark model for light hadrons, with measured anomalous magnetic moments for the proton and the neutron respectively $a_p = 1.79$ $a_n = -1.91$ give the estimation

$$1/3(4\mu_u - \mu_d) = 2.79$$
$$1/3(4\mu_d - \mu_u) = -1.91$$
$$a_u = 0.24, \quad a_d = 0.30$$

Nonlinear chiral quark model is also used to estimate the order of magnitude of the anomalous magnetic moments.

$$a_i \approx m_i^2 / \Lambda^2_{CSB} \approx 10\%$$
$$m_i^2 \approx 360 Mev$$
$$\Lambda^2_{CSB} \approx 1 Gev$$

$\Lambda_{CSB}$ is the Chiral symmetry breaking scale

## II- FEW DEFINITIONS

Baryon magnetic moments, longitudinal spin and transverse spin are defined as follows.

$$\vec{\mu}_N = \langle PS | \sum_{i,\bar{i}} \frac{Q_i}{2} \int dr^3 \vec{r} \times \bar{\psi}_i \vec{\gamma} \psi_i | PS \rangle + ...$$
$$\langle PS | \int dx^3 \psi_i^\dagger \vec{\Sigma} \psi_i | PS \rangle = 2\Delta_i \vec{S}$$
$$\langle PS | \int dx^3 \bar{\psi}_i \vec{\Sigma} \psi_i | PS \rangle = \vec{\delta}_i$$

To compute the baryon magnetic moment we use Gordon decomposition.
The convection current part:

$$\frac{x_i \mu_i}{2(1+x_i)}(\Delta_i - \frac{\delta_i}{x_i})$$

The spin current part:

$$\frac{x_i \mu_i}{2}(\Delta_i + \frac{\delta_i}{x_i})$$

Where $x_i = \frac{m_i}{\langle E_i \rangle}$ is the ratio of the constituent quark mass to the average kinetic energy of the quark in the baryon ground state. Adding antiquarks and denoting $\vec{\mu}_N = \langle P \uparrow | \vec{\mu}_N | P \uparrow \rangle$ and $\delta i = \delta_i - \delta_{\bar{i}}$ we get the upgraded KS formula.

$$\mu_N = \sum_{i=u,d,s} \mu_i W_i + ...,$$
$$\mu_i = \frac{Q_i}{2m_i}, \qquad (1)$$
$$2\frac{W_i}{x_i} = \frac{1}{(1+x_i)}(\Delta_i - \Delta_{\bar{i}} - \frac{\delta i}{x_i}) + (\Delta_i - \Delta_{\bar{i}} + \frac{\delta i}{x_i})$$

## III- TENSOR CHARGE AND ANOMALOUS MAGNETIC MOMENT CORRELATION.

KS formula (1) has an insufficiency. It leads to an absence of magnetism in the ultra-relativistic limit:

$$\mu_N = \sum_{i=u,d,s} \mu_i W_i + ..., \approx 0$$

This is due to (Lorentz-Fitzgerald contraction length)

$$x\mu_i \approx \frac{m_i}{\langle E_{i0} \rangle} \frac{1}{m_i} = \frac{1}{\langle E_{i0} \rangle} \approx 0$$

And to the quark masses being neglected in this regime:

$$\mu_N \mid_{mass} \sim (-\delta i + \delta i) = 0$$

The absence of magnetism in this limit suggests that KS formula does have a missing term. That term is the anomalous magnetic moment of the quark. In the following we introduce two different arguments to support our suggestion. First argument: KS formula describes a magnetic photon coupling to quarks being spinning point like objects.

Formula (1) is a relativistic formula which describes how a magnetic photon couples to quarks being spinning point like objects. It also says that this coupling is decreasing with energy due to the reduction factor . On the other hand we know from quantum mechanics that particles of definite energy and momentum are not localized. It then follows a possible current in the lagrangian of the form[1]

$$\frac{\partial_\alpha \bar{\psi} \sigma^{\alpha\beta} \psi_\beta}{m} \quad (2)$$

Perturbatively, for a photon to probe such a current, a quark should radiate a field ( gluon or goldstone boson or whatever ) at position $x$ and reabsorbed at a distant position $y$, once it interacts with the photon ( vertex interaction and not a self-energy interaction ).In this process the probing photon sees the quark as an extended object or rather an electric current circulating in the area of the extension .This is what we call "anomalous" magnetism. The correlation of the anomalous magnetic moment to the tensor charge is suggested by the structure of the current(2) which, as the mass term,

---

[1] Differentiation of the field is non zero only if the field has a spatial and/or temporal extension. Point like objects have a current without derivatives such as $\bar{\psi} \gamma_\alpha \psi$ for instance.

flips helicity. Adding quark anomalous magnetic moments of quarks to formula (1), this one generalizes to

$$\mu_N = \sum_{i=u,d,s} \mu_i W_i + ...,$$

$$2\frac{W_i}{x_i} = \frac{1}{(1+x_i)}(\Delta_i - \Delta_{\bar{i}} - \frac{\delta i}{x_i}) + (1+a_i)(\Delta_i - \Delta_{\bar{i}} + \frac{\delta i}{x_i}) \quad (3)$$

Second argument. Rearrange formula (1) as this.

$$2W_i = A_i(\Delta_i - \Delta_{\bar{i}}) + B_i(\delta_i - \delta_{\bar{i}})$$

$$\frac{A_i}{x_i} = 1 + \frac{1}{1+x_i}$$

$$B_i = 1 - \frac{1}{1+x_i}$$

$A_i$ and $B_i$ are not independent parameters ( being functions of only one common parameter $x_i$ ), but and are supposed to describe two independent spin contributions in a relativistic regime: Longitudinal and transverse polarizations.

To make $A_i$ and $B_i$ independent, we introduce an additional parameter $\varsigma_i$ in addition to $x_i$.

$$\frac{A_i}{x_i} = \frac{1}{1+x_i} + 1 + \varsigma_i$$

$$B_i = -\frac{1}{1+x_i} + 1 + \varsigma_i$$

We added the same contribution for both $A_i$ and $B_i$ as the anomalous magnetic moment does not depend on the polarization of particles.

Comparing with new KS leads to the identification of the parameter $\varsigma_i$ with the anomalous magnetic moment of quarks $a_i$

$$\varsigma_i = a_i$$

Through these arguments we understand that the introduction of anomalous magnetic moment is a necessary requirement of relativity. On the other it becomes also clear in this last approach, that the quark anomalous magnetic moment is correlated to the quark transversity. Such a correlation is manifest at the ultra relativistic limit at which $W_i$ functions in (3) take the form.

$$2W_i = a_i (\delta i)_{ultra}$$

where $(\delta i)_{ultra} = \frac{2}{3}\delta_i^{NR} = \frac{2}{3}\Delta_i^{NR}$. This limit makes it explicit that quark anomalous magnetic moments together with tensor charges dominate the ultra relativistic regime.

## IV- NUMERICAL ANALYSIS

Our results are as follows.

$$a_u \simeq a_d \simeq 0.38$$
$$\Delta_u - \Delta_{\bar{u}} \simeq 0.78$$
$$\Delta_d - \Delta_{\bar{d}} \simeq -0.39$$

We may use above axial magnetic densities $\Delta_u + \Delta_{\bar{u}} = 0.83$; $\Delta_d + \Delta_{\bar{d}} = -0.44$ to infer sea quark polarizations

$$\Delta_{\bar{u}} \simeq 0.03$$
$$\Delta_{\bar{d}} \simeq -0.05$$

For a complete analysis please refer to the paper of the author [7]